\documentclass[letterpaper,twocolumn,letter]{jpsj3}
\usepackage{txfonts}

\usepackage{xcolor}

\usepackage[final]{changes}

\usepackage{txfonts}
\usepackage{bm}

\definechangesauthor[name = Dai, color=red]{DA}
\definechangesauthor[color=green]{JF}

\newcommand{\DA}[1]{\added{#1}}

\title{Molten Salt Flux Liquid Transport Method for Ultra Clean Single Crystals UTe$_2$}

\author{
Dai~Aoki$^1$\thanks{E-mail: aoki@imr.tohoku.ac.jp}
}

\inst{%
$^1$IMR, Tohoku University, Oarai, Ibaraki 311-1313, Japan\\
}

\abst{%
Various single-crystal growth techniques have been presented for the unconventional superconductor UTe$_2$.
The molten salt flux liquid transport (MSFLT) method is employed to grow high-quality and large single crystals, exhibiting a high residual resistivity ratio ($\mbox{RRR}\sim200\mbox{--}800$).
However, the Te self-flux and chemical vapor transport (CVT) methods produce samples of lower quality.
The MSFLT method is a hybrid approach combining the molten salt flux (MSF) and CVT methods.
One significant advantage is that the materials gradually crystallize at a relatively low temperature, which is maintained throughout the main process.
This may be crucial for preventing U deficiency and obtaining high-quality large single crystals of UTe$_2$.
Many different single crystals obtained using different techniques have been characterized by resistivity and specific heat measurements.
The superconducting transition temperature $T_{\rm c}$ decreases with residual resistivity $\rho_0$,
followed by the Abrikosov-Gor'kov pair-breaking theory, 
and reaches $2.1\,{\rm K}$ for $\rho_0 \to 0$.
The residual $\gamma$-value of the specific heat for the highest quality sample was only $3\,{\%}$ of the normal-state $\gamma$-value.
The specific heat jump, $\Delta C_{\rm e}/(\gamma T_{\rm c})$ reached approximately $2.7$ for high-quality samples, indicating a strong-coupling superconductor.
Furthermore, the magnetic susceptibility for $H\parallel a$-axis in a high-quality single crystal does not show an up-turn behavior on cooling, which is consistent with the results of NMR Knight shift and $\mu$SR experiments.
}

\begin{document}
\maketitle

UTe$_2$ has attracted considerable attention because of its unusual superconducting properties ~\cite{Ran19,Aok19_UTe2,Aok22_UTe2_review}. 
The large upper critical field $H_{\rm c2}$ associated with the field-reentrant behavior
and multiple superconducting phases at pressure~\cite{Bra19,Aok20_UTe2,Tho21_PRB} as well as at high fields for $H\parallel b$-axis~\cite{Rosuel23,Sak23,Kin23_PRB}, demonstrate spin-triplet superconductivity owing to the spin and orbital degrees of freedom.
The determination of superconducting order parameters is an important task, and therefore, high-quality single crystals are required.
In the early stages of the discovery of superconductivity with insufficient sample quality, ``partially gapped superconductivity'' was proposed, owing to the large residual density of states in the specific heat measurements ~\cite{Ran19}.
However, recent high-quality single crystals have revealed an almost zero residual density of states~\cite{Aok22_UTe2_review},
and fully gapped superconductivity with spin-triplet $A_u$ state has been proposed~\cite{Mat23,Sue23}.
Single or double superconducting transition in the specific heat is also an important issue, because it is related to the breaking of time-reversal symmetry in the superconducting state and the components of the superconducting order parameters ~\cite{Hay21,Ish23,Aje23,Aza23_arXiv}.
In high-quality single crystals, a single superconducting transition of the specific heat is established at ambient pressure and zero field ~\cite{Tho21_PRB,Aok22_UTe2_review}. 

A breakthrough in high-quality single-crystal growth has recently been achieved ~\cite{Sak22}.
Owing to a new technique called the molten salt flux (MSF) method, the quality of single crystals UTe$_2$ has been improved significantly.
The superconducting transition temperature increased from $1.6$ to $2.1\,{\rm K}$,
and de Haas-van Alphen (dHvA) oscillations were observed for the first time~\cite{Aok22_UTe2_dHvA}.
The quasi-two-dimensional Fermi surfaces with heavy quasiparticles were directly detected using dHvA experiments.
These results were confirmed using different techniques by different groups ~\cite{Aok23_UTe2_dHvA,Eat23,Bro23}.

In this paper, we present our empirical results on different methods for the single-crystal growth of UTe$_2$ and their characterization.
The MSF method significantly improves the sample quality, resulting in a high residual resistivity ratio (RRR) of more than 100.
This technique is further improved by employing the so-called molten salt flux liquid transport (MSFLT) method,
which is a hybrid of the CVT and MSF methods.
Over 170 attempts at single-crystal growth using different techniques have yielded single crystals with different qualities.
These have been characterized using resistivity, specific heat, and magnetic susceptibility measurements.
Resistivity measurements were performed using the four-probe AC method with a current along $a$-axis.
Specific heat measurements were performed using a relaxation technique down to $0.34\,{\rm K}$.
The magnetic susceptibility was measured using a SQUID magnetometer down to $1.9\,{\rm K}$ for $H \parallel a$-axis.
Single-crystals of different qualities were also tested through dHvA experiments using a field-modulation technique.

The resistivity results reveal that $T_{\rm c}$ is dependent on the sample quality, with the maximum $T_{\rm c}=2.1\,{\rm K}$, and it follows the Abrikosov-Gor'kov pair-breaking theory.
The specific heat results show that $T_{\rm c}$ decreases with an increase in the residual density of states. The highest-quality sample showed an almost zero residual density of states and a large specific heat jump ($\Delta C_{\rm e}/(\gamma T_{\rm c}) \sim 2.7$), revealing strong coupling superconductivity.

The first attempt at single crystal growth involved a self-flux method using excess Te. 
The starting materials U and Te, with an atomic ratio of $\mbox{Te/U}=3.55$, were placed in an alumina crucible.
This was sealed in a tantalum tube in an arc furnace under Ar gas.
The tantalum tube was further sealed under vacuum in a quartz ampoule.
According to the U-Te binary alloy phase diagram~\cite{Mas90}, UTe$_2$ can grow above 950$^\circ$C.
The quartz ampoule was heated in an electrical furnace to 1050$^\circ$C and cooled to 955$^\circ$C at a cooling rate of $1.5\,^\circ{\rm C/hr}$. 
At this temperature, excess Te was removed via centrifugation.
Large single crystals of millimeter size were obtained.
Superconductivity is observed at $1.08\,{\rm K}$ defined by zero resistivity with a low residual resistivity ratio, $\mbox{RRR}=3.6$, despite the absence of bulk superconductivity found in the specific heat measurements.
The self-flux method is applicable to the growth of large single crystals of UTe$_2$; however, the sample quality was poor.

\begin{figure}[t]
\begin{center}
\includegraphics[width= 1\hsize]{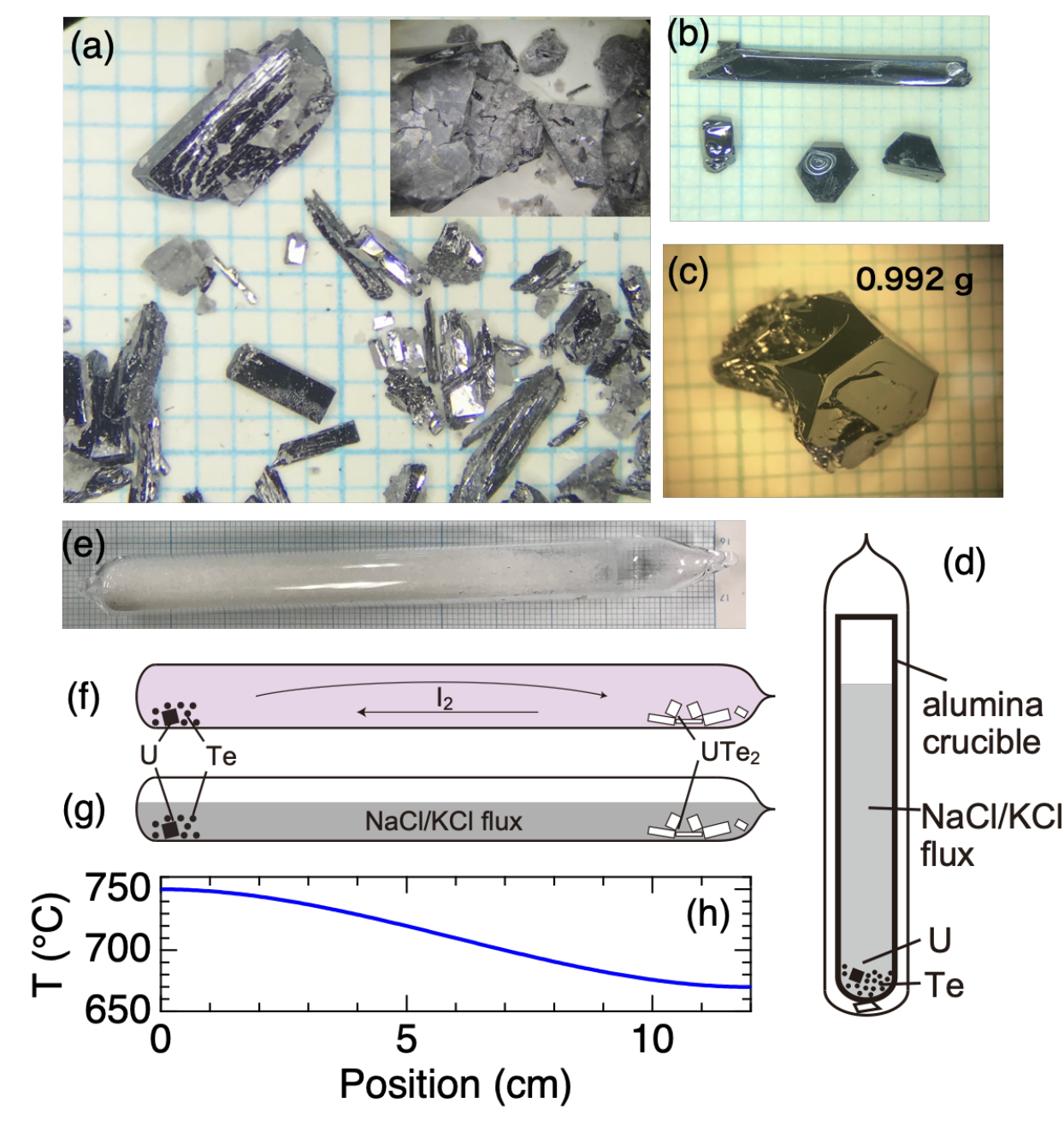}
\end{center}
\caption{(Color online) Photographs of single crystals UTe$_2$ grown by (a) MSFLT and (b) CVT. 
(c) A very large single crystal of 0.992 g can be grown by the CVT method. 
Schematics of quartz ampoules for the (d) MSF, (f) CVT, and (g) MSFLT methods, and \DA{(e) a photograph of a quartz ampoule for MSFLT method.}
(h) Temperature profile for the MSFLT method.}
\label{fig:photo}
\end{figure}

The CVT method with an off-stoichiometric amount of starting materials yielded better-quality samples. 
The starting materials in the ratio $\mbox{Te/U}\sim 1.5$ and iodine as the transport agent adjusted to a density of $\sim 3\,{\rm mg/cm^3}$
were put in a quartz ampoule sealed under vacuum.
The ampoule was then placed in a horizontal furnace and heated.
The temperature gradient with $\Delta T = 60\mbox{--}100\,^\circ{\rm C}$ was maintained for 10--14 days.
The starting materials were transported from the high-temperature side to the low-temperature side, where
single crystals were grown. 
Figures~\ref{fig:photo}(f) and \ref{fig:photo}(b) present a schematic of the CVT method and a photograph of single crystals, respectively.
By selecting optimal conditions, a large single crystal of $\sim 1\,{\rm g}$ can be grown, as shown in Fig. ~\ref{fig:photo}(c).

\begin{figure}[t]
\begin{center}
\includegraphics[width= 1\hsize]{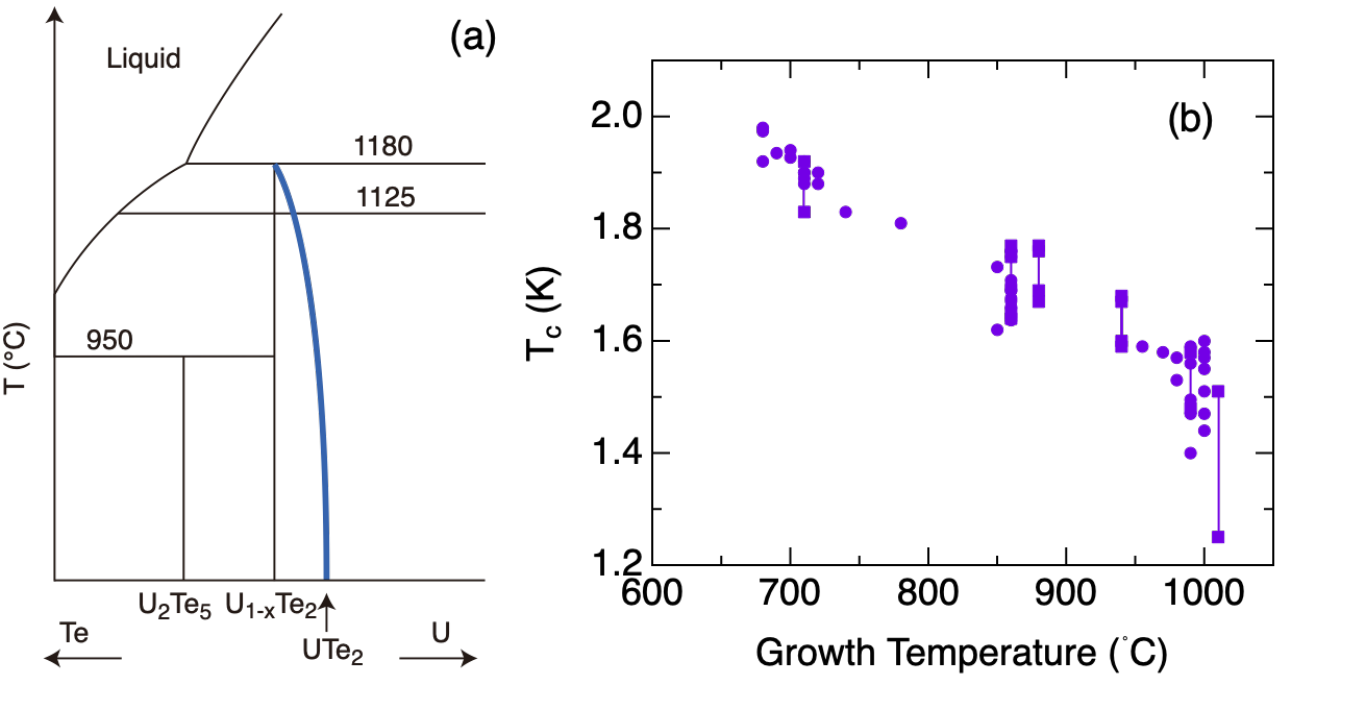}
\end{center}
\caption{(Color online) (a) Schematic Te-U phase diagram, which is slightly different from that in Ref.~\citen{Mas90}. 
(b) $T_{\rm c}$ vs. growth temperature at lower temperature side for CVT grown samples. Here, $T_{\rm c}$ is defined by the specific heat jump. The lines connecting two data points correspond to double transitions in specific heat.}
\label{fig:phase_Tc_growthT}
\end{figure}

The values of $T_{\rm c}$ depend on the growth temperature of the CVT method.
Figure~\ref{fig:phase_Tc_growthT}(b) shows the $T_{\rm c}$ determined by specific heat measurements as a function of the growth temperature on the lower-temperature side in 54 different samples grown using the CVT method.
$T_{\rm c}$ varies from $2$ to $\sim 1.4\,{\rm K}$, which
reveals that a lower growth temperature is preferable for a higher $T_{\rm c}$.
A similar trend was reported in Ref.~\citen{Ros22}. 

According to the single-crystal X-ray analysis~\cite{Hag22}, 
a low-quality sample without superconductivity is associated with a slight U deficiency of approximately $4\,\%$, 
whereas the superconducting sample exhibits a stoichiometric ratio of U to Te.
Thus, we propose a Te-U equilibrium phase diagram~\cite{Sak23_private}, as shown schematically in Fig. ~\ref{fig:phase_Tc_growthT}(a), which differs slightly from that reported in Ref.~\citen{Mas90}. 
Growth at high temperatures may yield small amounts of U deficiency, whereas a high-quality sample with stoichiometric U and Te is available through low-temperature growth.
Figure~\ref{fig:phase_Tc_growthT}(a) is also consistent with the fact that the Te self-flux method always produces a low-quality sample with U deficiency. 

Note that the starting composition for the CVT method must be off-stoichiometric and U-rich, specifically, $\mbox{Te/U} < 2$, to obtain high-quality single crystals with less U deficiency.
In fact, we obtained low-quality samples using the CVT method with stoichiometric starting materials, $\mbox{Te/U} = 2$, and a temperature gradient $950/850\,^\circ{\rm C}$, in which RRR is approximately 2.5. 
This was consistent with the results in previous studies~\cite{Ike06_UTe2,Cai20}.

The MSF method significantly improves the sample quality~\cite{Sak22}.
Single crystals can be grown at relatively low temperatures using this method, and the excess U in the molten salt acts as a reducing agent and suppresses the U vacancies.
A mixture of NaCl and KCl with an equi-atomic ratio was selected as flux, which has the eutectic melting point of $650\,^\circ{\rm C}$.
As shown in Fig.~\ref{fig:photo}(d), the starting materials and flux with an atomic ratio of U : Te : NaCl : KCl = 1 : 1.65 : 30 : 30 were placed in a long alumina crucible or directly into a quartz ampoule. 
The quartz ampoule was then evacuated with gentle heating at approximately $200\,^\circ{\rm C}$ to dehydrate the flux, and then sealed under vacuum.
The quartz ampoule was heated in an electrical furnace up to $950\,^\circ{\rm C}$ and maintained for one day. It was then slowly cooled to $650\,^\circ{\rm C}$ at a rate of $1\mbox{--}2\,^\circ{\rm c}/{\rm hr}$.
After dwelling at $650\,^\circ{\rm C}$ for 1 day, the furnace was switched off.
Single crystals were grown in salt at the bottom of the crucible.
When the ratio of the starting materials was selected as $\mbox{Te/U}=1.6\mbox{--}1.7$, the resistivity results showed
$\mbox{RRR}=50\mbox{--}220$ and $T_{\rm c}=2.00\mbox{--}2.08\,{\rm K}$, which are defined as zero resistivity.
The results of the specific heat yielded $T_{\rm c}=1.96\mbox{--}2.07\,{\rm K}$ and the residual $\gamma$-value scaled by the value in the normal state, $\gamma_0/\gamma_{\rm N}=0.034\mbox{--}0.16$.
A higher Te ratio, $\mbox{Te/U}=1.8$ produces a lower-quality sample with $\mbox{RRR}= 22$ and a lower $T_{\rm c}$ of $1.45\,{\rm K}$ at the specific heat.
Single crystals are not obtained with the atomic ratio of $\mbox{Te/U}=1.5\mbox{--}1.55$.

Most single crystals obtained by the MSF method are rather small and elongated along the $a$-axis, with a thin flat surface perpendicular to the $c$-axis or $[011]$ direction in reciprocal space.
Even if a thick single crystal is obtained, several halls may exist in the crystal, in which the layered crystals are piled up as in ``mille-feuille.''
This is a disadvantage of the MSF method compared to that in the CVT method.
Another problem is that the temperature gradually decreases during crystallization.
This may lead to samples of different quality \DA{as well as other impurity phases such as U$_7$Te$_{12}$ and U$_3$Te$_5$~\cite{Sak22}} in the crucible.

To obtain larger single crystals of higher quality, we developed a new technique called molten salt flux liquid transport (MSFLT) method.
The starting materials and salt flux were placed into a long quartz ampoule with a length of $120\mbox{--}190\,{\rm mm}$ and an inner diameter of $8\,{\rm mm}$.
The quartz ampoule was sealed under vacuum after gentle heating to dehydrate the salt ( Fig. ~\ref{fig:photo}(e) and (g)).
The ampoule was gradually heated in a horizontal furnace and maintained at a temperature gradient $750/670\,^\circ{\rm C}$ for 10--14 days, as shown in Fig. ~\ref{fig:photo}(h).

This is a hybrid approach comprising the CVT and MSF methods. 
Instead of vapor, liquid flux carries the starting material as a transport agent~\cite{Yan17}.
An advantage of the MSFLT method is that the growth temperature is fixed at a low temperature during the process, which might be favorable for high-quality single-crystal growth of UTe$_2$ because the lower growth temperature yields a higher quality sample, as shown in Fig. ~\ref{fig:phase_Tc_growthT}(b).
Another advantage of the MSFLT method is that relatively large, high-quality single crystals can be obtained, as shown in Fig.~\ref{fig:photo}(a).
This was because the growth conditions were stable at a fixed growth temperature, preventing nucleation in the ampoule during the process.
As the materials were transported, some impurities are expected to remain at the starting position as in the CVT method.

The crystal-growth conditions for the self-flux, CVT, MSF and MSFLT methods are summarized in Table~\ref{tab}, along with the results of characterization by specific heat and resistivity measurements.
\begin{fulltable}[tbh]
\newcommand{\N}{\phantom{0}}
\caption{Crystal growth conditions and results of characterizations in different samples obtained by self-flux, chemical vapor transport (CVT), molten salt flux (MSF) and molten salt flux liquid transport (MSFLT) methods, where Te/U is the atomic ratio of the starting material to Te and U. A mixture of NaCl and KCl with equi-atomic ratio was used for MSF and MSFLT.
The temperature decreased slowly in the self-flux and MSF methods, whereas a fixed temperature gradient was applied in the CVT and MSFLT methods.
RRR and $\gamma_0/\gamma_{\rm N}$ denote the residual resistivity ratio and residual $\gamma$-value scaled by the normal state $\gamma$-value, respectively.
$T_{\rm c}^{\rho=0}$ is $T_{\rm c}$ defined as the zero resistivity. $T_{\rm c}^{\rm C_p}$ is $T_{\rm c}$ defined by the specific heat jump.
$^\ast$No superconductivity was observed down to $1.7\,{\rm K}$ \DA{for sample \#S4}.
}
\label{tab}
\begin{center}
\begin{tabular}{llcccccccc}

\hline
Method	& Batch & Te/U & Flux & Temperature	($^\circ$C)&RRR	&$\gamma_0/\gamma_{\rm N}$	&$\Delta C_{\rm e}/\gamma T_{\rm c}$	&$T_{\rm c}^{\rho=0} ({\rm K})$	& $T_{\rm c}^{\rm C_p} ({\rm K})$\\
\hline
Self flux	&\#F2	&3.55	&Te			&$1050\!\to\!955\,(1.5^\circ{\rm C/h})$	&3.6	& --	&--		&1.08	&--	\\
CVT			&\#C5	&2.00	&--			&950\,/\,850							&2.5	& --	&--		&--		&--	\\
CVT			&\#C34	&1.50	&--			&1050\,/\,990							&14		&0.61\N	&1.29	&1.65	&1.47	\\
CVT			&\#C89	&1.40	&--			&780\,/\,680							&49		&0.13\N	&2.36	&2.01	&1.97	\\
MSF			&\#S4	&1.80	&NaCl/KCl	&$950\!\to\!650\,(1^\circ{\rm C/h})$	&22		&0.78\N	&0.58	&--$^\ast$	&1.45	\\
MSF			&\#S6	&1.65	&NaCl/KCl	&$950\!\to\!650\,(2^\circ{\rm C/h})$	&220	&0.046	&2.33	&2.06	&2.04	\\
MSFLT		&\#S56	&1.50	&NaCl/KCl	&750\,/\,670							&179	&0.124	&2.25	&2.06	&2.01	\\
MSFLT		&\#S37	&1.65	&NaCl/KCl	&750\,/\,670							&800	&0.034	&2.57	&2.09	&2.05	\\


\hline

\end{tabular}
\end{center}
\end{fulltable}

\begin{figure}[t]
\begin{center}
\includegraphics[width= 0.8\hsize]{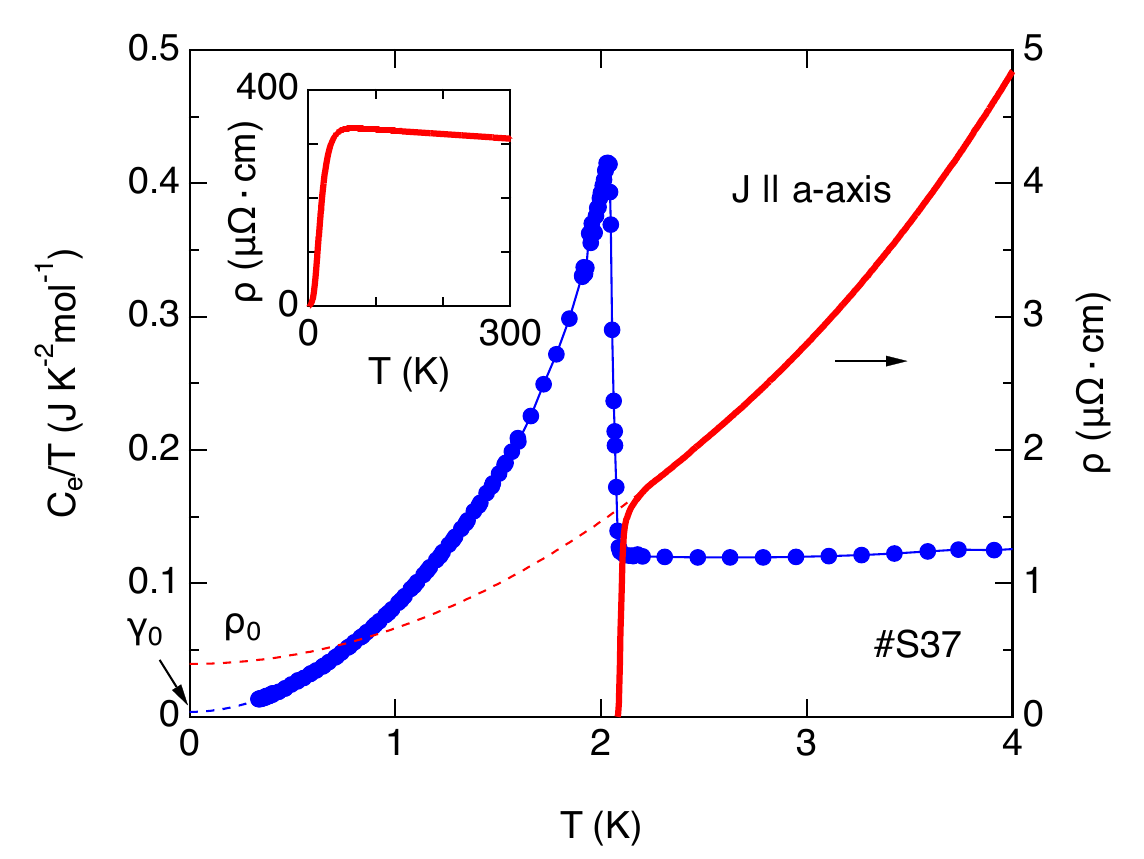}
\end{center}
\caption{(Color online) Electronic specific heat and resistivity at low temperatures for sample \#S37 with $\mbox{RRR}\sim 800$. A sharp single jump in the specific heat is observed at $T_{\rm c}^{\rm C_{\rm p}}=2.05\,{\rm K}$. The transition width is $\Delta T_{\rm c}\sim 0.05\,{\rm K}$. The residual $\gamma$-value scaled by the normal state $\gamma$-value is $\gamma_0/\gamma_{\rm N} = 0.034$. Zero resistivity is observed at $T_{\rm c}^{\rho=0}=2.09\,{\rm K}$. The onset of the superconducting transition is observed at approximately $2.2\,{\rm K}$. The inset shows the resistivity curve up to $300\,{\rm K}$. The dotted lines are results of fitting extrapolated to $0\,{\rm K}$.}
\label{fig:Cp_rho}
\end{figure}
Figure~\ref{fig:Cp_rho} shows the specific heat and resistivity measured at low temperatures for sample \#S37 grown using the MSFLT method.
The electronic specific heat, $C_{\rm e}$ was obtained by subtracting the phonon contribution.
A sharp single specific heat jump at $2.05\,{\rm K}$ was defined by the entropy balance with a transition width $\sim 0.05\,{\rm K}$. 
A large jump in $\Delta C_{\rm e}/\gamma T_{\rm c}=2.57$ reveals strong coupling superconductivity.
The data were fitted from $0.34$ to $0.65\,{\rm K}$ using $C_{\rm e}/T=\gamma_0 + B T^2$, which revealed
a very small residual $\gamma$-value of $\gamma_0/\gamma_{\rm N}=0.034$.
The resistivity of the current along the $a$-axis also shows a sharp drop with $T_{\rm c}=2.09\,{\rm K}$ defined by zero resistivity with an onset at approximately $2.2\,{\rm K}$.
The residual resistivity, $\rho_0$, obtained by the $T^2$ fitting below $3.5\,{\rm K}$ and RRR
are $0.40\,\mu\Omega\!\cdot\!{\rm cm}$ and 800, respectively.
Here, the resistivity at $300\,{\rm K}$ is scaled using a previously reported value ~\cite{Eo22}.

\begin{figure}[t]
\begin{center}
\includegraphics[width= 0.8\hsize]{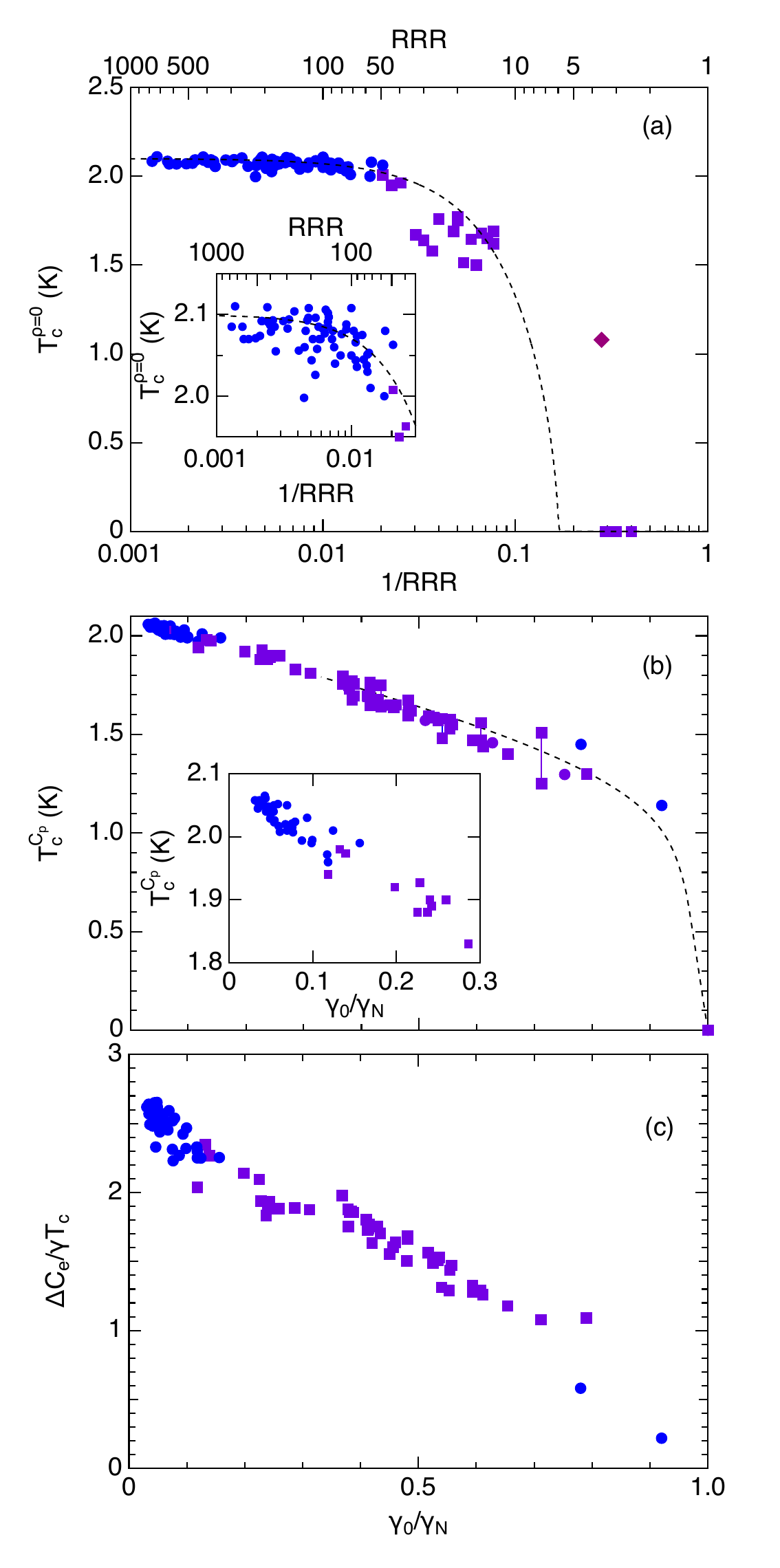}
\end{center}
\caption{(Color online) (a) $T_{\rm c}$ as a function of RRR or inverse RRR for different 91 samples grown by MSF/MSFLT (blue circle), CVT (purple square) and self-flux (red-purple diamond) methods. $T_{\rm c}$ is defined by zero resistivity. The inset shows the zoomed view at high RRR. The dotted line denotes the results of Abrikosov-Gor'kov pair breaking theory.
(b) $T_{\rm c}$ as a function of the residual $\gamma$-value scaled by the normal state $\gamma$-value for 95 samples.
The inset shows the zoomed view at low residual $\gamma$-values. The dotted line is a guide to the eyes.
(c) $\Delta C_{\rm e}/\gamma T_{\rm c}$ as a function of the scaled residual $\gamma$-value.
}
\label{fig:Tc_RRR_gamma0_DeltaC}
\end{figure}
Figure~\ref{fig:Tc_RRR_gamma0_DeltaC}(a) shows the relationship between $T_{\rm c}$ defined by zero resistivity, and the inverse RRR for 91 samples grown using the MSF, MSFLT, CVT, and self-flux methods.
$T_{\rm c}$ clearly depended on the sample quality and showed almost no change for $\mbox{RRR}\gtrsim 100$.
The maximum $T_{\rm c}$ is $2.1\,{\rm K}$ when $\rho_0 \to 0$.
High-quality samples were mainly obtained using the MSF or MSFLT methods.
The results approximately follow the Abrikosov-Gor'kov pair-breaking theory~\cite{Abr61}, as indicated by the dotted line.
%

The results of the specific heat measurements as bulk properties are shown in Fig. ~\ref{fig:Tc_RRR_gamma0_DeltaC}(b).
All measurements were performed down to $\sim 0.4\,{\rm K}$, and the residual $\gamma$-values were extracted by fitting $C_{\rm e}/T = \gamma_0 + BT^2$ below $0.65\,{\rm K}$.
A higher $T_{\rm c}$ indicates a lower residual $\gamma$-value, reaching $\gamma_0/\gamma_{\rm N}\sim 0.03$,
which implies that the residual $\gamma$-value is zero for an ideal crystal.
\DA{Note that the residual density of states and the residual thermal conductivity are theoretically discussed in Ref.~\citen{Min22}.}

The specific heat jump was also affected by the sample quality. 
Figure~\ref{fig:Tc_RRR_gamma0_DeltaC}(c) shows $\Delta C_{\rm e}/\gamma T_{\rm c}$ as a function of $\gamma_0/\gamma_{\rm N}$.
The value of $\Delta C_{\rm e}/\gamma T_{\rm c}$ increases with a decrease in the residual $\gamma$-value and reaches 2.7 for $\gamma_0/\gamma_{\rm N} \to 0$.

\begin{figure}[t]
\begin{center}
\includegraphics[width= 0.8\hsize]{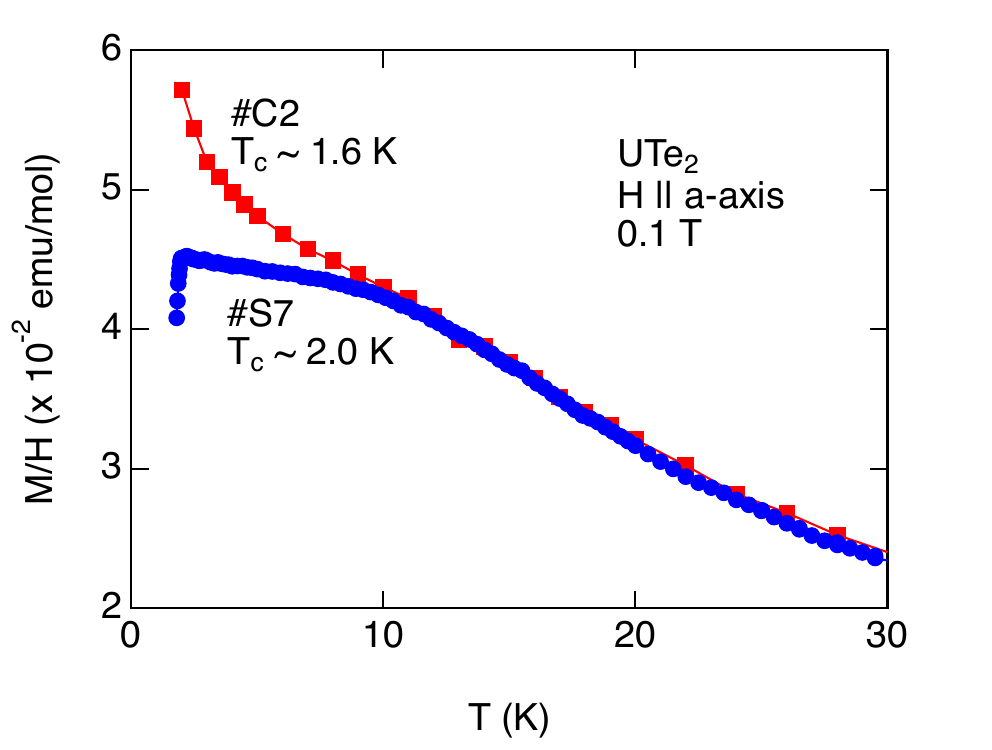}
\end{center}
\caption{(Color online) Magnetic susceptibilities at low temperatures for $H\parallel a$-axis at $0.1\,{\rm T}$ in two different quality samples grown by CVT ({\#}C2, $T_{\rm c}\sim 1.6\,{\rm K}$) and MSF ({\#}S7, $T_{\rm c}\sim 2.0\,{\rm K}$). The drop below $2\,{\rm K}$ for sample {\#}S7 is due to superconductivity.}
\label{fig:sus}
\end{figure}
The magnetic susceptibility for $H\parallel a$-axis, which corresponds to the easy-magnetization axis, also differs between high- and low-quality samples.
As shown in Fig.~\ref{fig:sus}, the low-quality sample with $T_{\rm c}=1.6\,{\rm K}$ showed an upturn upon cooling below $10\,{\rm K}$,
which had been analyzed by scaling to the ferromagnetic quantum criticality.
However, the high-quality sample with $T_{\rm c}=2\,{\rm K}$ shows no upturn below $10\,{\rm K}$,
revealing that the upturn behavior is not intrinsic, but due to disorder/defects or magnetic clusters.
These results are consistent with those obtained by the NMR Knight shift~\cite{Tok22} and $\mu$SR experiments~\cite{Aza23_PRB}.

\begin{figure}[t]
\begin{center}
\includegraphics[width= 0.8\hsize]{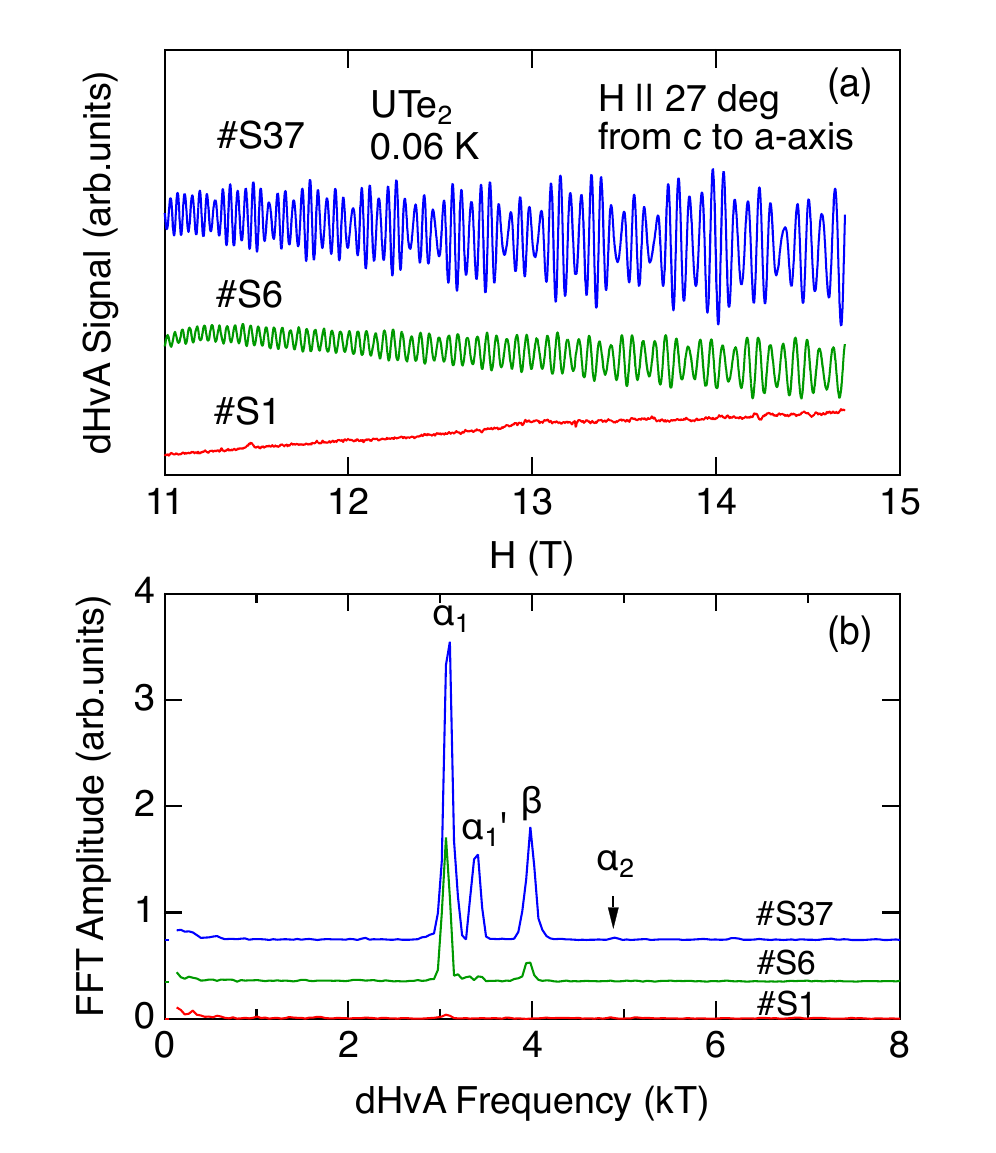}
\end{center}
\caption{(Color online) (a)dHvA oscillations at $0.06\,{\rm K}$ for the field direction tilted by $27$ deg from $c$ to $a$-axis in different samples \#S37, \#S6 and \#S1, and (b)the corresponding FFT spectra. \DA{The data are vertically shifted for clarity.} The values of RRR, $T_{\rm c}$ by specific heat, and $\gamma_0/\gamma_{\rm N}$ are
$\sim 800$, $2.05\,{\rm K}$ and $0.03$ for sample \#S37,
$\sim 220$, $2.04\,{\rm K}$ and $0.05$ for sample \#S6,
$\sim 100$, $1.99\,{\rm K}$ and $0.12$ for sample \#S1, respectively.
}
\label{fig:dHvA}
\end{figure}
The dHvA oscillations are highly sensitive to sample quality, particularly for frequencies with heavy effective masses.
Figure~\ref{fig:dHvA} shows the high-quality sample \#S37 with four fundamental dHvA branches, $\alpha_1$, $\alpha_1^\prime$, $\beta$ and $\alpha_2$, whereas the dHvA amplitudes are strongly suppressed in the lower-quality samples \#S6 and \#S1.
The Dingle temperatures for branch $\alpha_1$ \DA{with effective mass of $35\,m_0$} were $0.11$ and $0.15\,{\rm K}$ for samples \#S37 and \#S6, respectively.
The corresponding mean free paths were $1050$ and $830\,{\rm \AA}$, 
revealing a higher quality of sample \#S37.

In summary, different growth techniques for UTe$_2$ single crystals were presented.
High-quality single crystals were obtained using MSFLT, 
where $T_{\rm c}$ and the residual $\gamma$-value, $\gamma_0/\gamma_{\rm N}$, reached $2.1\,{\rm K}$ and $0.03$, respectively.
$T_{\rm c}$ approximately follows the Abrikosov-Gor'kov pair-breaking theory, and is related to the residual $\gamma$-values.
No upturn was observed in the susceptibility to $H \parallel a$-axis at low temperatures in the high-quality sample.
The MSFLT method is a hybrid approach between the MSF and CVT methods.
This technique, which is generally called the liquid transport method or horizontal flux method, can be applied to high-quality single-crystal growth of other materials~\cite{Che20,Cha23}
by leveraging the fixed growth temperature during this process.

\section*{Acknowledgements}
We thank Y. \={O}nuki, H. Sakai, J. Flouquet, G. Knebel, V. Mineev, Y. Haga, G. Lapertot, J. P. Brison, D. Braithwaite, Y. Homma, D. X. Li, A. Nakamura, Y. Shimizu and A. Miyake
for fruitful discussions and technical support.
This study was supported by KAKENHI (JP19H00646, JP20K20889, JP20H00130, JP20KK0061, JP22H04933).


\begin{thebibliography}{10}

\bibitem{Ran19}
S.~Ran, C.~Eckberg, Q.-P. Ding, Y.~Furukawa, T.~Metz, S.~R. Saha, I.-L. Liu,
  M.~Zic, H.~Kim, J.~Paglione, and N.~P. Butch: Science {\bfseries 365}, 684 (2019).

\bibitem{Aok19_UTe2}
D.~Aoki, A.~Nakamura, F.~Honda, D.~Li, Y.~Homma, Y.~Shimizu, Y.~J. Sato,
  G.~Knebel, J.-P. Brison, A.~Pourret, D.~Braithwaite, G.~Lapertot, Q.~Niu,
  M.~Vali\v{s}ka, H.~Harima, and J.~Flouquet: J. Phys. Soc. Jpn. {\bfseries 88}, 043702 (2019).

\bibitem{Aok22_UTe2_review}
D.~Aoki, J.~P. Brison, J.~Flouquet, K.~Ishida, G.~Knebel, Y.~Tokunaga, and
  Y.~Yanase: J. Phys.: Condens. Matter {\bfseries 34}, 243002 (2022).

\bibitem{Bra19}
D.~Braithwaite, M.~Vali{\v{s}}ka, G.~Knebel, G.~Lapertot, J.~P. Brison,
  A.~Pourret, M.~E. Zhitomirsky, J.~Flouquet, F.~Honda, and D.~Aoki: Commun.
  Phys. {\bfseries 2} 147, (2019).

\bibitem{Aok20_UTe2}
D.~Aoki, F.~Honda, G.~Knebel, D.~Braithwaite, A.~Nakamura, D.~Li, Y.~Homma,
  Y.~Shimizu, Y.~J. Sato, J.-P. Brison, and J.~Flouquet: J. Phys. Soc. Jpn.
  {\bfseries 89}, 053705 (2020).

\bibitem{Tho21_PRB}
S.~M. Thomas, C.~Stevens, F.~B. Santos, S.~S. Fender, E.~D. Bauer, F.~Ronning,
  J.~D. Thompson, A.~Huxley, and P.~F.~S. Rosa: 
Phys. Rev. B {\bfseries 104}, 224501 (2021).

\bibitem{Rosuel23}
A.~Rosuel, C.~Marcenat, G.~Knebel, T.~Klein, A.~Pourret, N.~Marquardt, Q.~Niu,
  S.~Rousseau, A.~Demuer, G.~Seyfarth, G.~Lapertot, D.~Aoki, D.~Braithwaite,
  J.~Flouquet, and J.~P. Brison: 
  Phys. Rev. X {\bfseries 13}, 011022 (2023).

\bibitem{Sak23}
H.~Sakai, Y.~Tokiwa, P.~Opletal, M.~Kimata, S.~Awaji, T.~Sasaki, D.~Aoki,
  S.~Kambe, Y.~Tokunaga, and Y.~Haga: 
  Phys. Rev. Lett. {\bfseries 130}, 196002 (2023).

\bibitem{Kin23_PRB}
K.~Kinjo, H.~Fujibayashi, S.~Kitagawa, K.~Ishida, Y.~Tokunaga, H.~Sakai, S.~Kambe, A.~Nakamura, Y.~Shimizu, Y.~Homma, D.~X.~Li, F.~Honda, D.~Aoki, K.~Hiraki, M.~Kimata, and T. ~Sasaki: Phys. Rev. B {\bfseries 107}, L060502 (2023).

\bibitem{Mat23}
H.~Matsumura, H.~Fujibayashi, K.~Kinjo, S.~Kitagawa, K.~Ishida, Y.~Tokunaga,
  H.~Sakai, S.~Kambe, A.~Nakamura, Y.~Shimizu, Y.~Homma, D.~Li, F.~Honda, and
  D.~Aoki: J .Phys. Soc. Jpn. {\bfseries 92}, 063701 (2023).

\bibitem{Sue23}
S.~Suetsugu, M.~Shimomura, M.~Kamimura, T.~Asaba, H.~Asaeda, Y.~Kosuge,
  Y.~Sekino, S.~Ikemori, Y.~Kasahara, Y.~Kohsaka, M.~Lee, Y.~Yanase, H.~Sakai,
  P.~Opletal, Y.~Tokiwa, Y.~Haga, and Y.~Matsuda: arXiv:2306.17549 .

\bibitem{Hay21}
I.~M. Hayes, D.~S. Wei, T.~Metz, J.~Zhang, Y.~S. Eo, S.~Ran, S.~R. Saha,
  J.~Collini, N.~P. Butch, D.~F. Agterberg, A.~Kapitulnik, and J.~Paglione:
  Science {\bfseries 373}, 797 (2021).

\bibitem{Ish23}
K.~Ishihara, M.~Roppongi, M.~Kobayashi, Y.~Mizukami, H.~Sakai, Y.~Haga,
  K.~Hashimoto, and T.~Shibauchi: Nat. Commun. {\bfseries 14}, 2966 (2023).

\bibitem{Aje23}
M.~Ajeesh, M.~Bordelon, C.~Girod, S.~Mishra, F.~Ronning, E.~Bauer, B.~Maiorov,
  J.~Thompson, P.~Rosa, and S.~Thomas: Phys. Rev. X {\bfseries 13}, 041019 (2023).

\bibitem{Aza23_arXiv}
N.~Azari, M.~Yakovlev, N.~Rye, S.~R. Dunsiger, S.~Sundar, M.~M. Bordelon, S.~M.
 Thomas, J.~D. Thompson, P.~F.~S. Rosa, and J.~E. Sonier: arXiv:2308.09773 .

\bibitem{Sak22}
H.~Sakai, P.~Opletal, Y.~Tokiwa, E.~Yamamoto, Y.~Tokunaga, S.~Kambe, and
  Y.~Haga: Phys. Rev. Mater. {\bfseries 6} 073401, (2022).

\bibitem{Aok22_UTe2_dHvA}
D.~Aoki, H.~Sakai, P.~Opletal, Y.~Tokiwa, J.~Ishizuka, Y.~Yanase, H.~Harima,
  A.~Nakamura, D.~Li, Y.~Homma, Y.~Shimizu, G.~Knebel, J.~Flouquet, and
  Y.~Haga: J .Phys. Soc. Jpn. {\bfseries 91} 083704, (2022).

\bibitem{Aok23_UTe2_dHvA}
D.~Aoki, I.~Sheikin, A.~McCollam, J.~Ishizuka, Y.~Yanase, G.~Lapertot,
  J.~Flouquet, and G.~Knebel: J. Phys. Soc. Jpn. {\bfseries 92} 065002, (2023).

\bibitem{Eat23}
A.~G. Eaton, T.~I. Weinberger, N.~J.~M. Popiel, Z.~Wu, A.~J. Hickey, A.~Cabala,
  J.~Posp\'{i}\v{s}il, J.~Prokle\v{s}ka, T.~Haidamak, G.~Bastien, P.~Opletal,
  H.~Sakai, Y.~Haga, R.~Nowell, S.~M. Benjamin, V.~Sechovsk\'{y}, G.~G.
  Lonzarich, F.~M. Grosche, and M.~Vali\v{s}ka:
  Nat. Commun. {\bfseries 15}, 223 (2024).

\bibitem{Bro23}
C.~Broyles, Z.~Rehfuss, H.~Siddiquee, K.~Zheng, Y.~Le, M.~Nikolo, D.~Graf,
  J.~Singleton, and S.~Ran: arXiv:2303.09050 .

\bibitem{Mas90}
T.~B. Massalski, H.~Okamoto, P.~R. Subramanian, and L.~Kacprzak: {\em Binary
  Alloy Phase Diagrams (2nd ed.)} (ASM International, 1990).

\bibitem{Ros22}
P.~F.~S. Rosa, A.~Weiland, S.~S. Fender, B.~L. Scott, F.~Ronning, J.~D.
  Thompson, E.~D. Bauer, and S.~M. Thomas: 
  Communications Materials {\bfseries 3}, 33 (2022).

\bibitem{Hag22}
Y.~Haga, P.~Opletal, Y.~Tokiwa, E.~Yamamoto, Y.~Tokunaga, S.~Kambe, and
  H.~Sakai: J. Phys.: Condens. Matter {\bfseries 34} 175601, (2022).

\bibitem{Sak23_private}
H.~Sakai: private communication.

\bibitem{Ike06_UTe2}
S.~Ikeda, H.~Sakai, D.~Aoki, Y.~Homma, E.~Yamamoto, A.~Nakamura, Y.~Shiokawa,
  Y.~Haga, and Y.~{\=O}nuki: 
  J. Phys. Soc. Jpn. Suppl {\bfseries 75}, 116 (2006).

\bibitem{Cai20}
L.~P. Cairns, C.~R. Stevens, C.~D. O'Neill, and A.~Huxley: J. Phys.: Condens.
  Matter {\bfseries 32}, 415602 (2020).

\bibitem{Yan17}
J.-Q. Yan, B.~C. Sales, M.~A. Susner, and M.~A. {McGuire}: Phys. Rev. Mater.
  {\bfseries 1}, 023402 (2017).

\bibitem{Eo22}
Y.~S. Eo, S.~Liu, S.~R. Saha, H.~Kim, S.~Ran, J.~A. Horn, H.~Hodovanets,
  J.~Collini, T.~Metz, W.~T. Fuhrman, A.~H. Nevidomskyy, J.~D. Denlinger, N.~P.
  Butch, M.~S. Fuhrer, L.~A. Wray, and J.~Paglione: Phys. Rev. B {\bfseries
  106}, L060505 (2022).

\bibitem{Abr61}
A. A. Abrikosov and L. P. Gor'kov: Sov. Phys. JETP {\bfseries 1243}, 12 (1961).


\bibitem{Min22}
V. P. Mineev:
J .Phys. Soc. Jpn. {\bfseries 91}, 074601 (2022).


\bibitem{Tok22}
Y. Tokunaga, H. Sakai, S. Kambe, Y. Haga, Y. Tokiwa, P. Opletal, H. Fujibayashi, K. Kinjo, S. Kitagawa, K. Ishida, A. Nakamura, Y. Shimizu, Y. Homma, D. Li, F. Honda, and D. Aoki: 
J .Phys. Soc. Jpn. {\bfseries 91}, 023707 (2022).

\bibitem{Aza23_PRB}
N.~Azari, M.~R. Goeks, M.~Yakovlev, M.~Abedi, S.~R. Dunsiger, S.~M. Thomas,
  J.~D. Thompson, P.~F.~S. Rosa, and J.~E. Sonier: 
  Phys. Rev. B {\bfseries 108}, L081103 (2023).

\bibitem{Che20}
J.~Chen, M.~B. Gam\.{z}a, J.~Banda, K.~Murphy, J.~Tarrant, M.~Brando, and F.~M.
  Grosche: Phys. Rev. Lett. {\bfseries 125}, 237002 (2020).

\bibitem{Cha23}
G.~Chajewski, D.~Szyma\'{n}ski, M.~Daszkiewicz, and D.~Kaczorowski: Mater.
  Horiz.  (2023) DOI: 10.1039/d3mh01351k.

\end{thebibliography}

\end{document}